\documentclass[aps,prd,twocolumn,superscriptaddress]{revtex4}
\pdfoutput=1

\usepackage{graphicx}         % fuer das Einbinden von Grafiken
\usepackage{amssymb}
\usepackage{amsmath}
\usepackage{amsthm}
\usepackage{slashed}
\usepackage[colorlinks]{hyperref}
\usepackage{color,rotating}
\usepackage{bbm}
\usepackage{array}
\usepackage{pstricks}
\usepackage{pstricks-add}
\usepackage{colonequals}

\definecolor{darkgreen}{rgb}{0,0.6,0}
\definecolor{gray}{rgb}{.7,.7,.7}

\DeclareMathAlphabet{\EuRoman}{U}{eur}{m}{n}
\SetMathAlphabet{\EuRoman}{bold}{U}{eur}{b}{n}

 \graphicspath{{./figures/}}
\def\di{\displaystyle}

\def\bg{\begin{eqnarray}\begin{array}{rcl}\displaystyle}
\def\eg{\end{array} &\di    &\di   \end{eqnarray}}
\def\bm#1{\begin{eqnarray}\begin{array}{#1}\di}
\def\bmo#1{\begin{eqnarray*}\begin{array}{#1}\di}
\def\bml#1#2{\begin{eqnarray}\begin{array}{#1}\label{#2}\di}
\def\bgo{\begin{eqnarray*}\begin{array}{rcl}\displaystyle}
\def\ego{\end{array} &\di    &\di \nonumber  \end{eqnarray*}}

\def\btensor#1#2{\renew\left#1\begin{array}{#2}\di}
\def\brtensor#1#2#3{\ren#3\left#1\begin{array}{#2}}
\def\botensor#1#2{\renew\left#1\begin{array}{#2}}
\def\etensor#1{\end{array}\right#1}

\def\d{{d}}

\def\s0#1#2{\mbox{\small{$ \frac{#1}{#2} $}}}
\def\0#1#2{\frac{#1}{#2}}

\def\s{\sigma}

\def\ren#1{\renewcommand{\arraystretch}{#1}}

\def\renew{\renewcommand{\arraystretch}{1}}
\allowdisplaybreaks

\begin{document}

\title{Global solutions of functional fixed point equations via pseudo-spectral methods}

% LIST_OF_AUTHORS.TEX                 5/18/01            
%
\author{J. Borchardt}
%\email[]{julia.borchardt@uni-jena.de}
\affiliation{Theoretisch-Physikalisches Institut, Universit\"at Jena,
Max-Wien-Platz 1, 07743 Jena, Germany}
\author{B. Knorr}
%\email[]{benjamin.knorr@uni-jena.de}
\affiliation{Theoretisch-Physikalisches Institut, Universit\"at Jena,
Max-Wien-Platz 1, 07743 Jena, Germany}

%end                                                                          

\begin{abstract}

We apply pseudo-spectral methods to construct global solutions of functional renormalisation group equations in field space to
high accuracy. For this, we introduce a basis to resolve both finite as well as asymptotic regions of effective
potentials. Our approach is benchmarked using the critical behaviour of the scalar $O(1)$ model, providing results
for the global fixed point potential as well as leading critical exponents and their respective global eigenfunctions.
We provide new results for (1) multi-critical $O(1)$ models in fractional dimensions, (2) the three-dimensional Gross-Neveu
model at both small and large $N$, and (3) the scalar-tensor model, also in three dimensions.

\end{abstract}

\maketitle

\section{Introduction}

Many interesting physical phenomena are characterised by strong coupling. Among them, there are
very fundamental problems as for example confinement in QCD or potentially the quantisation of
gravity. Conventional methods from quantum field theory, such as perturbation theory,
are not applicable in such cases. In recent years, an increasingly successful method has been
the functional renormalisation group (FRG). It is based on the
Wilsonian idea of integrating out modes momentum-shell by momentum-shell. In the last two decades,
the FRG, based on the formulation by Wetterich \cite{Wetterich:1992yh}, was successfully applied to a range of topics,
including scalar field theories \cite{Tetradis:1993ts,Berges:2002,Delamotte:2007pf,Morris:1997xj,Bervillier:2007rc,Litim:2010tt,Benitez:2011xx,
Codello:2012ec,Percacci:2014tfa,Codello:2014yfa},
fermionic systems \cite{Metzner:2011cw,Scherer:2010sv,Gies:2010st,Braun:2010tt,Braun:2011pp,Gies:2014xha,Boettcher:2013kia},
critical phenomena \cite{Kopietz:2010zz,Janssen:2012pq,Gies:2009da,Scherer:2013pda,Litim:2002cf,Benitez:2009xg,Jakubczyk:2014isa},
gauge theories \cite{Reuter:1993kw,Gies:2006wv,Pawlowski:2005xe,Litim:1998qi,Braun:2011iz,Tripolt:2013zfa,Braun:2014ata,Mitter:2014wpa}
and quantum gravity \cite{Reuter:1996cp,Niedermaier:2006wt,Percacci:2007sz,Manrique:2009uh,Benedetti:2010nr,Benedetti:2012dx,Eichhorn:2015bna,Demmel:2014hla,
Christiansen:2012rx,Falls:2013bv,Christiansen:2014raa,Folkerts:2011jz,Eichhorn:2011pc,Harst:2011zx,Dona:2013qba}.
From a technical perspective, the generic outcome of applying the FRG to a given model is a coupled system of non-linear
(integro-)differential equations of complex structure.
This is due to the full field- and momentum-dependent propagator entering the Wetterich equation.
There exist only a few cases where the full equations can be solved analytically.
In all other cases one has to consider the system within some truncation, retaining only a manageable number of operators.
Even then, the equations are rarely analytically solvable, e.g. in a large-$N$ or mean-field approximation.
However, if one seeks a solution without such approximations, numerical methods appear indispensable.
Various methods are used depending on the special structure of the system of equations.
All these methods aim at a numerically accurate solution that fulfils the equations to high precision.

In this work, we advocate the use of pseudo-spectral methods,
especially referring to Chebyshev polynomials as basis functions as a versatile tool for FRG equations.
The fall-off of the corresponding expansion coefficients provides a measure for the accuracy of the approximate solution.
Pseudo-spectral methods are a well-suited, fast means to treat a wide range
of different problems: ordinary and partial differential equations as well as eigenvalue problems \cite{Boyd:ChebyFourier}.
These are all problems relevant both in the FRG formalism and physics in general.

Here, we focus on calculating global solutions of physical systems,
especially referring to ordinary differential equations (ODEs).
In particular, we cover the basics of the pseudo-spectral concepts, and ultimately
apply them to several interesting problems arising from FRG applications.
Some of them, as the Ising model in three dimensions,
have already been the focus of many detailed studies in the past, making them a perfect testing ground
for the methods presented here. 
It is convenient to start with these models to discuss some mathematical and technical details
of the equations and their implementation. In this way, one can also easily compare to known results.
In order to emphasise that the range of applications of the methods presented here is very large,
we present a variety of applications and extract several new results. This includes multi-critical phenomena
in non-integer dimensions, and the three-dimensional Gross-Neveu model.
For the scalar-tensor model proposed recently in \cite{Percacci:2015wwa},
we gain new insights which cannot be obtained within local expansions.
Let us also point out that the methods presented here are heavily used in other contexts \cite{Boyd:ChebyFourier,Robson:1993}, as e.g. finding solutions to
Einstein’s equation \cite{Ansorg:2003br,Macedo:2014bfa}.
First applications to FRG problems have been given in \cite{Litim:2003kf,Fischer:2004uk,Gneiting:2005}. Our approach is particularly suited for global aspects and also resolves
asymptotic behaviour in a controlled way.
Recently, it has been used to globally resolve the supersymmetric analogue of the Wilson-Fisher fixed point \cite{Heilmann:2014iga}.

This work is organised as follows: in \autoref{sec:PSM}, we present the basic ideas of pseudo-spectral methods,
focusing on the properties essential for the subsequent discussion. Afterwards, \autoref{sec:IMP} sheds light on the specific
application of the methods to a given problem, while \autoref{sec:FRG} is a short overview on the FRG.
We then turn our attention to the $O(1)$ model in \autoref{sec:ON}, first studying the
Wilson-Fisher fixed point in three dimensions. Furthermore, we extend known results on multi-critical fixed points in non-integer
dimensions. Consequently, \autoref{sec:GNmodel} treats the Gross-Neveu model, first in the large flavor number limit, then considering
finite flavor numbers. Finally, \autoref{sec:STG} discusses a scalar-tensor model which couples a scalar field non-minimally to gravity.

The numerical results were obtained with code written in C++, including the libraries BOOST \cite{boost} for handling arbitrary precision and
Eigen \cite{eigen} for dealing with linear algebra. To analyse and present the data, Mathematica 10 \cite{Mathematica10} was employed.

\section{Pseudo-spectral methods}\label{sec:PSM}

Pseudo-spectral methods aim to represent a function via a suitable expansion.
Suitable means that the expansion shall be accurate and easily treatable.
In the present context, all derivatives needed should be easily computable to high accuracy
and they and the function itself should be evaluable at arbitrary points.
Natural candidates for such an expansion are the classical orthogonal polynomials,
which have some convenient properties regarding convergence, evaluation and taking derivatives.
In this paper, we will focus on the Chebyshev polynomials of the first kind, which are
defined by
\begin{equation}\label{eq:Tndef}
 T_n(\text{cos}(x)) = \text{cos}(n x), \quad n \in \mathbbm{N}_0 \, ,
\end{equation}
and their cousins, the rational Chebyshev polynomials \cite{Boyd:ratCheb},
\begin{equation}
 R_n(x) = T_n\left( \frac{x-L}{x+L} \right) \, .
\end{equation}

Here, $L>0$ is an arbitrary parameter, encoding the precise compactification in $x$.
The reason for this choice is that they have superior convergence properties as compared to
Legendre polynomials or Chebyshev polynomials of the second kind. Both Hermite and Laguerre
polynomials are ill-suited for our problems for the following reason: they are defined on
unbounded intervals, and increasing the interpolation order changes the asymptotic behaviour. For
the problems usually encountered, the asymptotic behaviour is fixed and thus the convergence
properties of Hermite and Laguerre polynomials is difficult to control. One possibility is to use 
the Hermite or Laguerre functions, which decay exponentially, but again, the convergence properties are
often worse compared to those of rational Chebyshev polynomials.

In the following, we will collect some important properties of Chebyshev polynomials. Similar relations
hold for their rational counterparts. We will only state results, for deeper information and proofs,
consider e.g. \cite{Boyd:ChebyFourier}.
First, a fast way of evaluating an expansion in Chebyshev polynomials,
\begin{equation}
 f(x) = \sum_{i=0}^N a_i T_i(x) \, ,
\end{equation}
at an arbitrary point $x$ is given by the Clenshaw algorithm, which is based on the 3-term recurrence
relation
\begin{equation}
 T_{n+1}(x) = 2x T_n(x) - T_{n-1}(x) \, .
\end{equation}

With this, the function $f(x)$ is readily evaluated via the recursive algorithm
\begin{align}
 b_{N+2} = b_{N+1} &= 0 \, , \notag \\
 b_i &= a_i + 2x b_{i+1} - b_{i+2} \, , \notag \\
 f(x) &= a_0 + x b_1 - b_2 \, .
\end{align}

Second, the derivative of $f(x)$ can again be expanded in a sum of Chebyshev polynomials of degree $N-1$.
The expansion coefficients, call them $a_i'$, are given recursively by
\begin{align}
 a_{N-1}' &= 2 N a_N \, , \notag \\
 a_{N-2}' &= 2 (N-1) a_{N-1} \, , \notag \\
 a_i' &= 2(i+1) a_{i+1} + a_{i+2}' \, .
\end{align}

Both the Clenshaw and the derivative algorithm are high-performance algorithms and numerically stable due to their recursive nature.

The third, but most important property of the Chebyshev polynomials is the exceptional convergence of the expansion coefficients, 
which is based to their relation to Fourier series as indicated by \eqref{eq:Tndef}. To be more precise, let us first define the
algebraic index of convergence as the largest number $k$ for which
\begin{equation}
 \lim_{n \rightarrow \infty} |a_n| n^k < \infty \, .
\end{equation}

So-called exponential convergence is achieved if the coefficients $a_n$ decrease faster than $1/n^k$ for any $k$.
For example, the Chebyshev expansion of a Lipschitz continuous function is always converging exponentially.
One can further differentiate exponential convergence into supergeometric, geometric and subgeometric convergence,
but these details shall not matter here. On a log-log plot, algebraic convergence manifests itself in a straight line,
whereas exponential convergence is indicated by a bending downwards. Note that the definition of convergence is
an asymptotic one, and might set in only when including a large number of coefficients.

By Darboux's principle, convergence properties of a series is related to the singularity structure of the function to be interpolated.
This includes poles, branch cuts, fractional powers, discontinuities in the function or in any of its derivatives etc. in
the complex plane.
Important in this context is the convergence domain of a Chebyshev series. It is given by the interior of an ellipse
whose foci lie at $x=\pm 1$. This is in contrast to a Taylor series (or more generally a Laurent series), whose
domain of convergence is a disc around the expansion point.

We shall close this mini-review by recalling the Chebyshev truncation theorem, which gives an upper bound for the error made
in truncating a Chebyshev series. The error is given by the sum of the absolute value of the neglected coefficients.
A useful rule of thumb is that this error is of the order of the last retained coefficient for exponential convergence,
and of the order of the number of coefficients retained times the last coefficient for algebraic convergence.

\section{Implementation}\label{sec:IMP}

In this section, we want to give more details on how to apply pseudo-spectral methods by considering a generic ODE of one function in
one variable. Without loss of generality we restrict to the domain $x \in \mathbbm R_+$. If the domain is $\mathbbm R$, then
we apply the subsequent ideas to both $\mathbbm R_+$ and $\mathbbm R_-$.

Let $\mathcal L$ be a (not necessarily linear) integro-differential operator and consider the problem
\begin{equation}\label{eq:generic_ode}
 \mathcal L \left[ f(x) \right] = 0 \, .
\end{equation}

We want to decompose the function $f(x)$ into a series of Chebyshev polynomials. To gain maximum efficiency,
first the domain of $f(x)$ is decomposed into M parts. We will restrict the discussion here
to $M=2$, that is we decompose the domain into $[0,x_0]$ and $[x_0,\infty]$. In the first domain,
the function is interpolated via a standard Chebyshev series, whereas in the second part, a rational Chebyshev series
is used. Thus
\begin{equation}
 f(x) = \begin{cases} \sum\limits_{i=0}^{N_c} c_i T_i(\frac{2x}{x_0}-1) , \, x \leq x_0 \, , \\
 f_{\infty}(x)\sum\limits_{i=0}^{N_r} r_i R_i(x-x_0) , \, x \geq x_0 \, , \end{cases}
\end{equation}
where $f_{\infty}(x)$ is the leading term of the asymptotic behaviour of $f(x)$ for $x \to \infty$,
which can be easily determined analytically in many cases. For FRG equations, the asymptotics is
typically determined by dimensional scaling properties.
This ansatz can be inserted into \eqref{eq:generic_ode}. To solve for the coefficients, it is useful to apply the
collocation method. It consists of evaluating the equation on a certain set of collocation points and solving
the resulting algebraic set of equations, e.g. by a Newton-Raphson method. The key to high accuracy is the choice of these collocation points. It turns out
that the best choice is to employ either the nodes or the extrema of Chebyshev polynomials \footnote{For some problems, it is
also useful to define left or right Radau-type collocations, which include either the left or the right end point,
respectively.}.
Additionally, one must match the function value
as well as derivatives of both expansions at the intermediate point $x_0$ to achieve smoothness. If the differential
equation is of order $p$, then $p-1$ derivatives have to be matched.

It should be noted that in the above decomposition, one has two free parameters: the matching point $x_0$ as well
as $L$, which encodes the compactification of the semi-infinite domain. There is no intrinsic rule how to
choose them. We found it reasonable to choose $x_0$ large enough to include the essential physics, e.g. the 
vacuum expectation value in a scalar field theory. The influence of $L$ is usually small in this case, as then the rational Chebyshev
expansion essentially only interpolates the asymptotic behaviour. Either way, if the expansion converges, it will
converge for any choice of the parameters.

The above ideas can be generalised in two ways. Firstly, one can trivially apply these methods to a system of functions,
i.e. a coupled system of ODEs. An example of this will be given later, and can also be found
in \cite{Heilmann:2014iga}. Secondly, via a tensor product, the generalisation to PDEs is possible. This will
be addressed in future work.

\section{The functional renormalisation group}\label{sec:FRG}

A very efficient means to store the full quantum information of a quantum field theory is the so-called effective action $\Gamma$,
which is defined as the Legendre transform of the Schwinger functional.
There are numerous possibilities to compute $\Gamma$, one is given by the FRG.
Instead of $\Gamma$, the effective average action $\Gamma_k$ is considered,
which smoothly interpolates between a microscopic theory $\Gamma_{k=\Lambda} = S_\text{cl}$, where $\Lambda$ is an ultraviolet cut-off,
and the full quantum theory $\Gamma_{k=0} = \Gamma$.
Following Wilson's idea, quantum fluctuations at momentum scale $p^2 \simeq k^2$ are successively integrated out during this evolution.
This process is described by an exact FRG equation, the Wetterich equation \cite{Wetterich:1992yh},
\begin{equation} \label{eq:Wetterich}
 k \partial_k \Gamma_k = \frac{1}{2} \mathrm{STr}\left[ \left(\Gamma^{(2)}_k + R_k \right)^{-1} (k \partial_k R_k) \right] \, ,
\end{equation}
where $\Gamma^{(2)}_k$ denotes the second functional derivative of $\Gamma_k$ with respect to the fields
and the super-trace $\mathrm{STr}$ stands for a summation over discrete indices, integration over continuous indices
and an additional minus sign for Grassmann-valued fields, i.e. fermions.
The functional $R_k$ is a regulator, which ensures both infrared as well as ultraviolet finiteness. Detailed information on the FRG can be found in e.g.
\cite{Berges:2002,Pawlowski:2005xe,Gies:2006wv,Berges:2012ty}.

In most cases, the exact functional integro-differential equation \eqref{eq:Wetterich} can only be solved by choosing a certain truncation for the effective average action.
A class of common truncations is the derivative expansion, which takes derivative interactions up to a given order into account.
For many cases, such a systematic expansion yields results comparable to those obtained by other methods, e.g. lattice studies.

\section{\texorpdfstring{$O(1)$}{O(1)} model near criticality}\label{sec:ON}

This section is devoted to a detailed study of various properties of the $O(1)$ model. 
Our ansatz for the effective average action reads
\begin{equation}
 \Gamma_k[\sigma] = \int \mathrm{d}^d x \left\{ \frac{1}{2} Z_k(\sigma^2) (\partial_\mu \sigma)^2 + U_k(\sigma^2) \right\},
\end{equation}
which contains the effective potential $U_k(\sigma^2)$ and a wave function renormalisation $Z_k(\sigma^2)$.
The operators are chosen such that the $\mathbb{Z}^2$ symmetry of the scalar field $\sigma$ is preserved.
In first order derivative expansion, also called local potential approximation (LPA),
one neglects the running and the field dependence of the wave function renormalisation, $Z_k(\sigma^2) \equiv 1$.
By contrast, within next-to-leading order in the derivative expansion (NLO), the full flow of the wave function renormalisation is taken into account.
As a compromise between LPA and NLO, one often considers a field-independent but scale-dependent wave function renormalisation
$Z_k \equiv Z_k(\sigma_0^2)$, usually called LPA'. Here, $\sigma_0$ is typically chosen to be the vacuum expectation value of the scalar field.

Here, we first study the Wilson-Fisher fixed point in three dimensions in LPA and LPA'. 
Although there exists a full analytic solution for the large-$N$ case, we will confine ourselves to the case of $N=1$.
A large-$N$ study is given below for the Gross-Neveu model in \autoref{sec:GNmodel}.
Subsequently, we look for multi-critical fixed points in dimensions
$2<d<3$ \cite{Codello:2014yfa}.

\subsection{Wilson-Fisher fixed point in LPA and LPA'}

\begin{figure}
\includegraphics[width=\columnwidth]{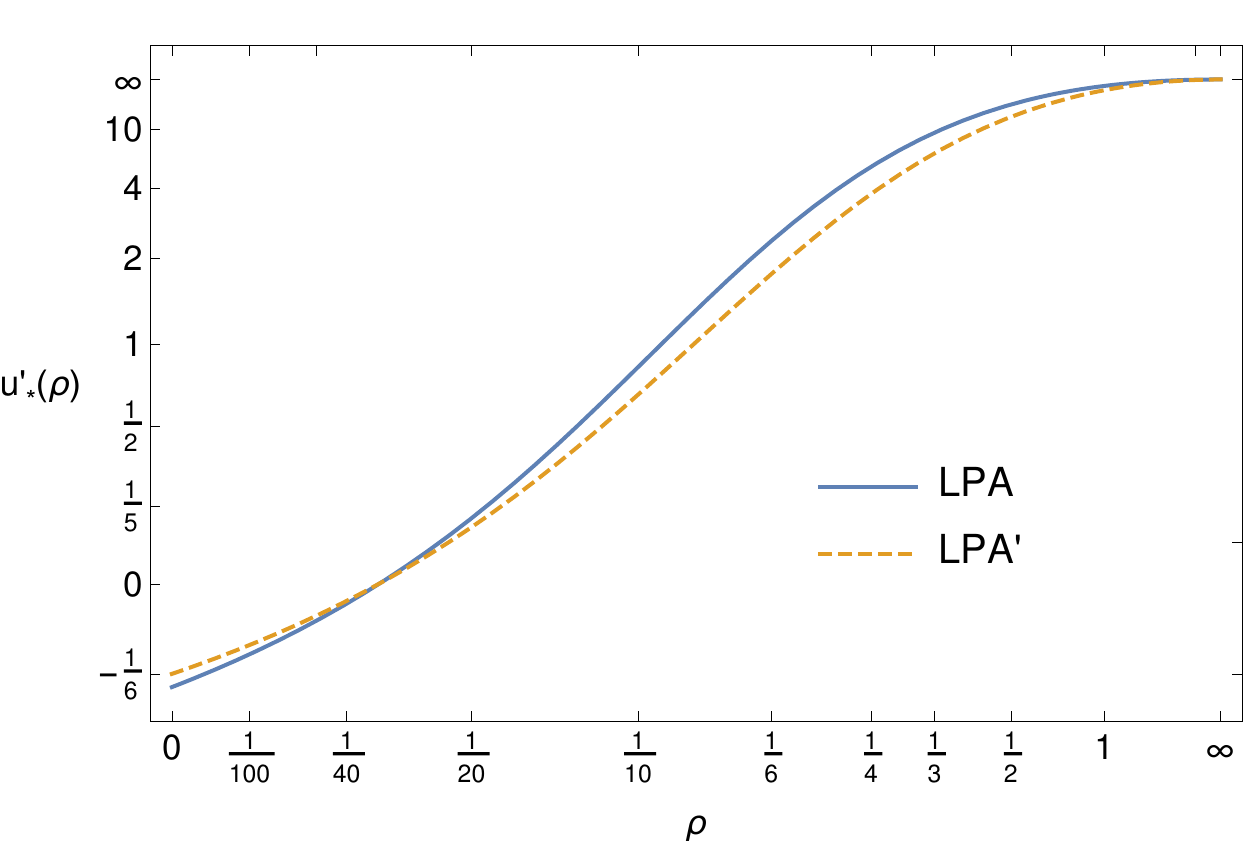}
\caption{Derivative of the effective potential of the Wilson-Fisher fixed point in LPA and LPA'.}
\label{fig:WF_comparison}
\end{figure}
First, let us study the well-known Wilson-Fisher fixed point of the $O(1)$ model in three dimensions. In arbitrary dimension
$d$, the fixed point equation is given by \cite{Tetradis:1993ts}
\begin{align}\label{eq:WF}
 &(-2+\eta)u'(\rho)+(d-2+\eta)\rho u''(\rho) \notag \\ 
 &- \frac{4 v_d}{d}\left( 1-\frac{\eta}{d+2} \right) \frac{3u''(\rho)+2\rho u'''(\rho)}{(1+u'(\rho)+2\rho u''(\rho))^2} = 0 \, ,
\end{align}
where $u(\rho) = k^{-d}U(\sigma^2)$ is the dimensionless effective potential as a function of the dimensionless invariant
$\rho = Z_k k^{2-d}\sigma^2/2$, $v_d^{-1} = 2^{d+1}\pi^{d/2}\Gamma(d/2)$ 
and $\eta=-k\partial_k \ln Z_k$ is the anomalous dimension. The latter is given by
\footnote{In fact, this is the anomalous dimension corresponding to the here absent ($N-1$) Goldstone modes. For some reason,
this gives superior results also in the case $N=1$. This also gave rise to some confusion in the literature. Strictly speaking,
here we consider the LPA' flow of an $O(N)$ model in the limit $N\to 1$.}
\begin{equation}\label{eq:WFeta}
 \eta = \frac{16 v_d}{d}\frac{\rho_0 u''(\rho_0)^2}{(1+2\rho_0 u''(\rho_0))^2} \, ,
\end{equation}
$\rho_0$ being the vacuum expectation value (vev). 
In these equations the optimised regulator is employed \cite{Litim:2001up}.
For aspects of optimisation, see also \cite{Litim:2000ci}.
As \eqref{eq:WF} does not depend on the potential itself, all our calculations will
involve its derivative instead, i.e. $f(x)\to u'(\rho)$.

In the following, we will compare the solution to \eqref{eq:WF} in LPA, i.e. with $\eta=0$,
and in LPA', where we include the anomalous dimension, but no field-dependent wave-function renormalisation.
As numerical parameters, $x_0=3/10$ and $L=1$ were chosen, and we used {\tt float128} which gives twice as many figures as the conventional
double data type. From \eqref{eq:WF}, one infers the asymptotic behaviour $u'_\infty(\rho) = \rho^{(2-\eta)/(d-2+\eta)}$.

\autoref{fig:WF_comparison} displays the derivative of the effective potential. One can see that the inclusion of the anomalous
dimension has a quantitative influence for intermediate values of $\rho$. For the vev, we get
\begin{align}
 \rho_0^\text{LPA} &= 0.030647942408697774953 \, , \notag \\
 \rho_0^\text{LPA'} &= 0.030592776234779436405 \, .
\end{align}

\begin{figure*}
\includegraphics[width=\textwidth]{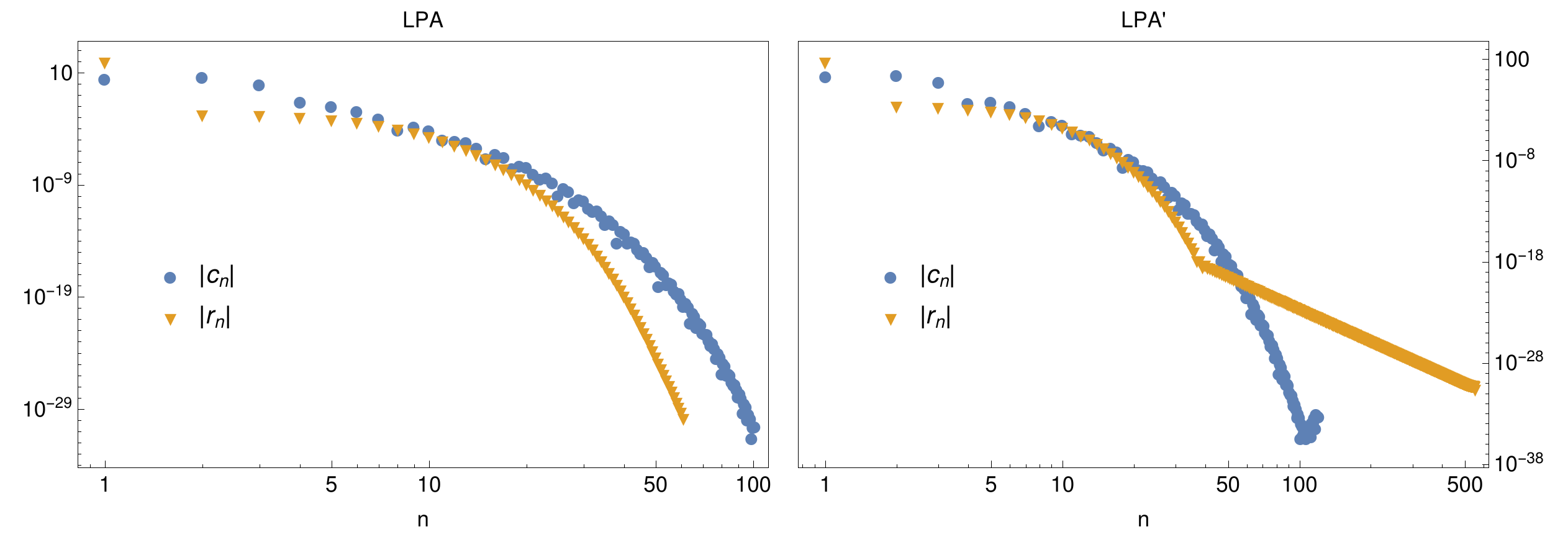}
\caption{Decay of coefficients of Chebyshev and rational Chebyshev expansion of the derivative of the Wilson-Fisher fixed point potential, LPA on the left and LPA' on the right.
Notice the algebraic decay in the latter case in the rational Chebyshev region, which needs a factor of 10 as many
coefficients as in the case of exponential convergence in LPA to achieve the same order of accuracy.}
\label{fig:WF_LogLog_coeffs}
\end{figure*}
In LPA', we found the anomalous dimension to be
\begin{equation}
 \eta = 0.044272337370315035214 \, .
\end{equation}

It may seem ridiculous to present that many figures, but they illustrate the power of pseudo-spectral methods. Our viewpoint here
is to solve a truncated problem (numerically) exactly, and any numbers given here are to be understood as
the solution to the truncated problem.

We find that our values match very well with earlier results, e.g. given in \cite{Morris:1997xj,Berges:2002,Codello:2012ec},
and \cite{Adams:1995cv} where fixed point quantities were calculated via full potential flows.
In particular, we can reproduce all digits of the high precision results given in \cite{Bervillier:2007rc}, where an LPA truncation
was used (notice the rescaling of $\rho$ by $8\pi^2$ compared to our conventions).

In comparison to results obtained from High-Temperature expansions and Monte-Carlo simulations ($\eta = 0.036$) \cite{Pelissetto:2000ek},
our value has a systematic error which is to be expected in the LPA' truncation.
Within the FRG, an accuracy competitive with other methods has been reached using the BMW approximation technique \cite{Benitez:2009xg,Benitez:2011xx}.

\autoref{fig:WF_LogLog_coeffs} shows the coefficients of the expansion of $u'(\rho)$ in a Chebyshev series and
of $u'(\rho)/\rho^{\frac{2-\eta}{1+\eta}}$ in a rational Chebyshev series.
In LPA, one perfectly sees exponential convergence in both the Chebyshev as well as the rational Chebyshev series. On the other hand,
as soon as we include the anomalous dimension, we find asymptotically only algebraic convergence in the rational Chebyshev case. This
behaviour is indeed expected by the asymptotic behaviour of the potential, as it rises with a fractional power. Furthermore, one can
also see that this problem is irrelevant for all practical purposes, as the algebraic convergence only sets in at about $10^{-18}$,
up to that point one still observes exponential convergence. This emphasises the fact that any statement about convergence is really
an asymptotic one, and one cannot predict where this behaviour sets in. As a final comment on this, note also the number of coefficients
needed to gain a certain accuracy: in case of exponential convergence, one needs very few coefficients to get an adequate result, but as soon
as there are singularities of any kind, one needs a large number of coefficients to further increase the accuracy.

Another point of interest is that no additional condition has to be imposed in solving \eqref{eq:WF}, in particular no boundary condition
or the like. Only the asymptotic behaviour $u'(\rho) \propto \rho^{\frac{2-\eta}{1+\eta}}$ and a sufficiently good initial guess is needed.
This may seem unexpected for a differential equation of second order, but can be understood along the lines of \cite{Dietz:2012ic}. Indeed, analysing
the situation, one can see that at $\rho=0$, the order of the differential equation decreases by one, which fixes one condition, and the same
is true at $\rho=\infty$.

As a further case for the method, let us expand the solution in LPA into a Taylor series around vanishing field as well as a Laurent series
around $\rho=\infty$ and compare whether the relations between coefficients obtained by plugging in such an ansatz into the fixed point equation
are satisfied. Around vanishing field, with $u'_*(\rho) = \sum a_i \rho^i$, one obtains the well-known relations (see e.g. \cite{Litim:2002cf})
\begin{align}
 a_1 &= -4\pi^2 a_0 (1+a_0)^2 \notag \, , \\
 a_2 &= \frac{12}{5}\pi^4 a_0 (1+a_0)^3(1+13 a_0) \notag \, , \\
 a_3 &= -\frac{288}{7} \pi^6 a_0^2 (1+a_0)^4(1+7a_0) \notag \, , \\
 a_4 &= \frac{32}{7} \pi^8 a_0^2 (1+a_0)^5 (2+a_0(121+623a_0)) \, ,
\end{align}
etc. Inserting our solution, one finds that the absolute error in these coefficients are
$(<10^{-30},2\times10^{-23}, 2\times10^{-19}, 7\times10^{-16})$. For the expansion around infinity, one finds that
\begin{equation}
 u'_*(\rho) = A \rho^2 - \frac{1}{75 A \pi^2 \rho^3} + \mathcal O(\rho^{-5}) \, .
\end{equation}

Expanding our solution, the coefficients of $\rho^1, \rho^0, \rho^{-1}, \rho^{-2}$ (which should vanish in the exact solution)
are $(-4\times10^{-27},3\times10^{-24},-8\times10^{-22},10^{-19})$, and the relation between the leading and the first sub-leading
coefficient is fulfilled to an absolute accuracy of $10^{-17}$. For completeness, let us give the values of $a_0$ and $A$ both in LPA and LPA':
\begin{align}
 a_0^\text{LPA} &= -0.18606424947031443565 \, , \notag \\
 a_0^\text{LPA'} &= -0.16574071049155738982 \, , \notag \\
 A^\text{LPA} &= 84.182303273336100651 \, , \notag \\
 A^\text{LPA'} &= 50.323366981670544177 \, .
\end{align}
These results match with \cite{Litim:2002cf} and \cite{Bridle:2013sra} where local expansions and the shooting method were employed.
This underlines that we can trust the global solution and that we can
relate to earlier results.

\begin{figure}
\includegraphics[width=\columnwidth]{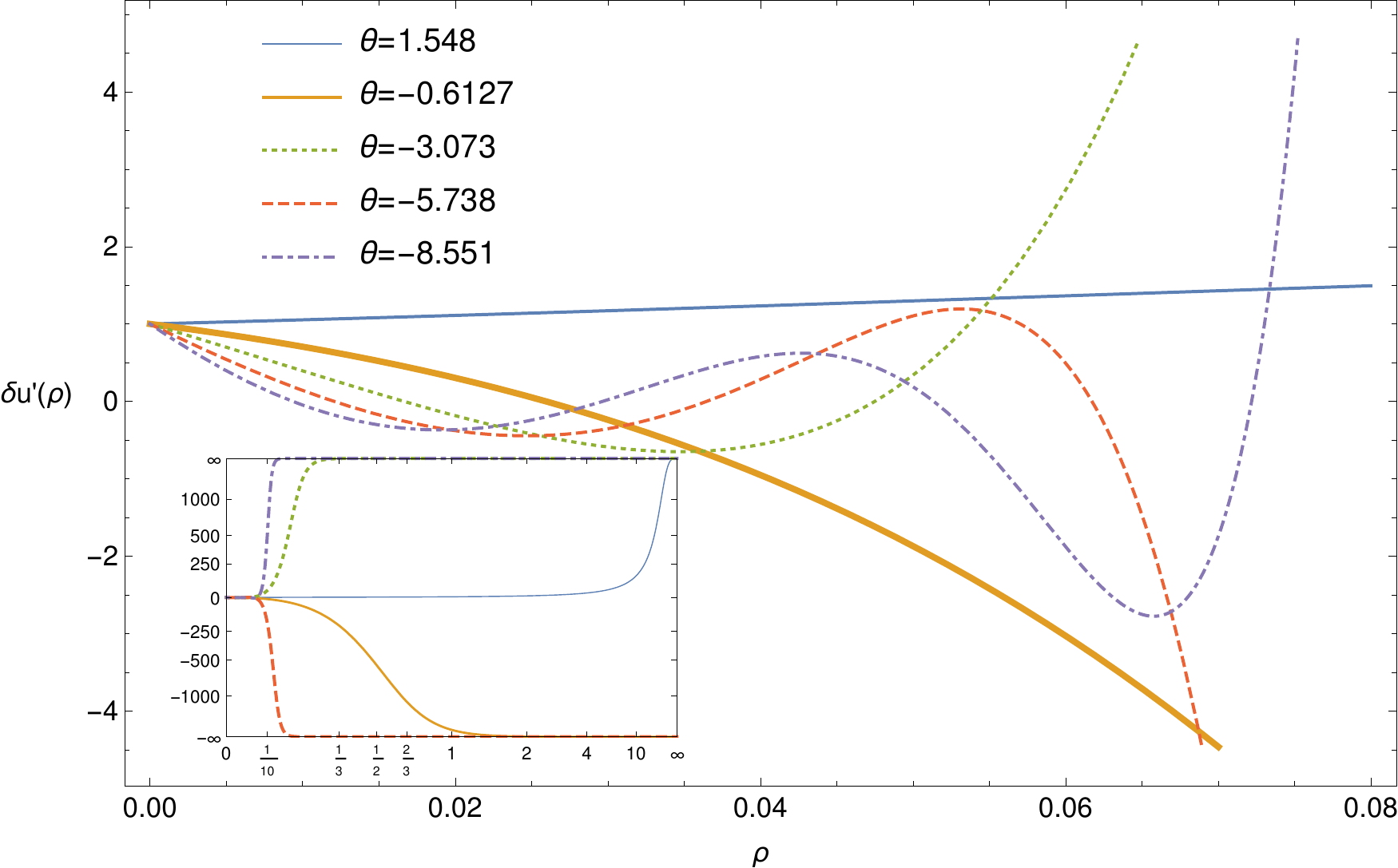}
\caption{Eigenperturbations of the Wilson-Fisher fixed point, normalised to one at $\rho=0$.}
\label{fig:WF_LPAprime_eigenfuncs}
\end{figure}
\begin{figure*}
\includegraphics[width=\textwidth]{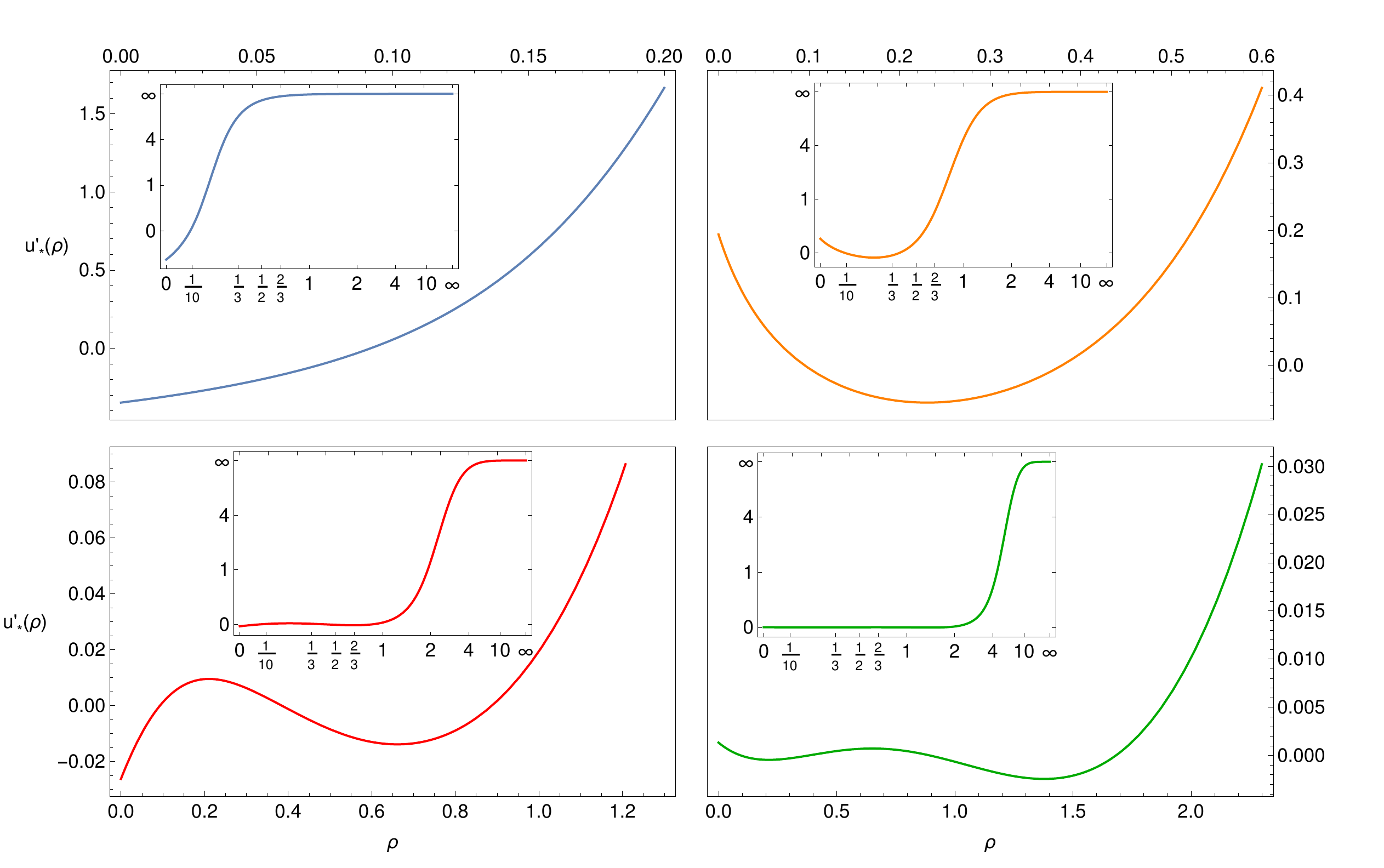}
\caption{First derivative of multi-critical fixed point potentials exhibiting two minima regarded as function of the dimensionless scalar field
(blue, Wilson-Fisher potential), three minima (yellow), 
four minima (red), five minima (green). The small insets depict the global behaviour of the solutions.}
\label{fig:ON_multicritFP}
\end{figure*}
Let us now turn our attention to the critical exponents of the Wilson-Fisher fixed point. They are defined as minus the eigenvalues
of the linearisation of the perturbed fixed point equation.
Again, a global approach to the solution of the perturbed equation is used. \autoref{fig:WF_LPAprime_eigenfuncs}
shows the eigenfunctions corresponding to the five highest eigenvalues, where the anomalous dimension has been taken into account.
As for the potential itself, any precision can be achieved in the eigenfunctions and critical exponents.
The critical exponents match with earlier results, e.g. given in \cite{Berges:2002}.
Having said that, let us emphasise again that the largest error arises from the systematic errors of the
derivative expansion to order LPA/LPA'.
If we compare with Monte-Carlo results \cite{Pelissetto:2000ek}, we find a deviation of about $2.5\%$ for the first and $27\%$ for the second critical exponent.
Especially the error of the second critical exponents is to be expected from the low order of the derivative expansion used, see \cite{Morris:1997xj}.

As shown in this example, the error is dominated by truncating the effective average action and not by numerical errors.
For this reason, from now on we will only give a few
relevant digits, bearing in mind that in principle we could calculate as many digits as needed.

\subsection{Multi-critical fixed points for \texorpdfstring{$2<d<3$}{2<d<3}}

It is worthwhile to have a closer look at fractional dimensions $2<d<3$.
The fixed point structure is getting richer for decreasing dimension and, therefore, it is interesting
to investigate the interpolation between the two fixed points in $d=3$, the Gaussian and the Wilson-Fisher fixed point, and the infinite number of fixed points in $d=2$.
In \cite{Codello:2012sc,Codello:2014yfa} the existence and properties of multi-critical fixed points in dependence on $d$ and $N$ are investigated.
These results are used to give an RG proof of the Mermin-Wagner-Hohenberg theorem \cite{Mermin:1966fe,Hohenberg:1967zz,gelfert:2001,Coleman:1973ci}.

As a test case and as has been done in \cite{Codello:2012ec} we restrict ourselves to the Ising universality class $N=1$ here.
We emphasise that the following investigations can straightforwardly be applied to arbitrary flavor numbers $N$
if the set of fixed point solutions is still discrete.
In \cite{Codello:2012ec}, a sequence of critical dimensions $d_{c,i}$ where the next multi-critical fixed point potential $u_i(\rho)$ emerges was suggested.
These are those dimensions where new operators $\rho^i$ become relevant for $d<d_{c,i}$.
Concentrating in the following on $d=2.4$ as an example, 
we find three more multi-critical fixed points FP${}_{i\in \{3,4,5\}}$ besides the Wilson-Fisher (FP${}_{2}$) and the Gaussian (FP${}_{1}$) fixed point.
The index $i\geq2$ counts the minima of the corresponding fixed point potential counted in the dimensionless scalar field.
Our results around $d=2.4$ confirm the predicted value $d_{c,6}=\frac{12}{5}$.

For our calculations we have employed \eqref{eq:WF} and \eqref{eq:WFeta} within the LPA' truncation.
The anomalous dimension is again evaluated at the global minimum of the potential which is in the following cases the outermost minimum.
In \autoref{fig:ON_multicritFP} the first derivative of the multi-critical fixed point potentials are shown.
As the values of the anomalous dimension of the multi-critical fixed points FP${}_{i\geq3}$ are small compared to the one of the Wilson-Fisher fixed point,
the convergence of their coefficients is exponential within the used precision of {\tt float128}. 
Therefore, the deviation from the exact solution can be estimated to be below $10^{-30}$.

In \autoref{tab:multicrit} the anomalous dimensions and the largest critical exponents calculated by pseudo-spectral methods are given.
Our results are in good agreement with \cite{Codello:2012ec,Codello:2012sc,Codello:2014yfa}.
Additionally, the results for the Wilson-Fisher fixed point in $d=2.4$ can be related to earlier works \cite{LeGuillou:1987ph,El-Showk:2013nia,Bhanot:1985rx},
where the $\varepsilon$-expansion and lattice simulations were applied.
As already noted in \cite{Codello:2014yfa}, the highest relevant critical exponent for the fixed points $i\geq$3 
is close to the mean field value $2$ at the corresponding critical dimension.
The other relevant exponents are smaller.
\begin{table}
\begin{tabular}{|c|c|c|}
\hline
 \multicolumn{3}{|c|}{WF-FP}\\ \hline
 $\eta$ & relev. exp. & irrelev. exp. \\ \hline
 $0.1390$ & $1.1441$ & $-0.7919$ \\ 
 & & $-3.1129$  \\ 
 & & $-5.6370$ \\ \hline
%  $\eta$ & $0.13898542503191918663837762808303$  \\ \hline
%  \multicolumn{2}{|c|}{$\theta_1=-1.1441$} \\ 
%  \multicolumn{2}{|c|}{$\theta_2=0.7919$, $\theta_3=3.1129$} \\ \hline
 \multicolumn{3}{|c|}{multi-critical FP$_{i=3}$}\\ \hline
  $\eta$ & relev. exp. & irrelev. exp. \\ \hline
 $0.01598$ & $1.9629$ & $-0.5108$ \\ 
 & $0.8416$ & $-2.0698$  \\ 
 & & $-3.8140$ \\ \hline
%  $\eta$ & $0.015980814998049000946816436272124$\\ \hline
%  \multicolumn{2}{|c|}{$\theta_1=-1.9629$, $\theta_2=-0.8416$} \\ 
%  \multicolumn{2}{|c|}{$\theta_3=0.5108$, $\theta_4=2.0698$, $\theta_5=3.8140} \\ \hline
 \multicolumn{3}{|c|}{multi-critical FP$_{i=4}$}\\ \hline
   $\eta$ & relev. exp. & irrelev. exp. \\ \hline
 $0.001753$ & $1.9969$ & $-0.3138$ \\ 
 & $1.4615$ & $-1.3968$  \\ 
 & $0.6726$ &  \\ \hline
%  $\eta$ &  $0.0017531058618824110679399889631646$\\ \hline
%  \multicolumn{2}{|c|}{$\theta_1=-1.9969$, $\theta_2=-1.4615$, $\theta_3=-0.6726$} \\ 
%  \multicolumn{2}{|c|}{$\theta_4=0.3138$, $\theta_5=1.3968$} \\ \hline
 \multicolumn{3}{|c|}{multi-critical FP$_{i=5}$}\\ \hline
    $\eta$ & relev. exp. & irrelev. exp. \\ \hline
 $8.2715\times10^{-5}$ & $1.9999$ & $-0.1655$ \\ 
 & $1.5973$ &  $-0.9297$ \\ 
 & $1.1243$ &  \\ 
 & $0.5414$ &  \\ \hline
%  $\eta$ &  $8.2714889832072538531878442328149\times10^{-5}$\\ \hline
%  \multicolumn{2}{|c|}{$\theta_1=-1.9999$, $\theta_2=-1.5973$, $\theta_3=-1.1243$, $\theta_4=-0.5414$} \\ 
%  \multicolumn{2}{|c|}{$\theta_5=0.1655$} \\ \hline
\end{tabular}
\caption{Anomalous dimensions and highest critical exponents of all scaling solutions in $d=2.4$.}
\label{tab:multicrit}
\end{table}

The sequence of critical dimensions predicts that a new fixed point potential with six minima (regarded as function of the dimensionless scalar field) emerges exactly at $d=2.4$.
As the Wilson-Fisher fixed point probably does not exist in $d=4$ but exists in all dimensions $2<d<4$ we find this fixed point for all dimensions $d<2.4$.
In fact, we are able to determine a global solution for $d=2.399$ where the non-asymptotic behaviour is realised on very smalls scales $|u'(\rho\leq\rho_0)|\sim 10^{-6}$
and $\eta = 2.3446\times 10^{-10}$.

\section{Gross-Neveu-model in \texorpdfstring{$d=3$}{d=3}}
\label{sec:GNmodel}

In this section we extend our studies to the partially bosonised Gross-Neveu model in $d=3$ dimensions. 
Loosely speaking, this is a generalisation of the $O(1)$ model including $N$ fermionic degrees of freedom.
It has applications in condensed matter physics and serves as a toy model for asymptotic safety scenarios.
A detailed analysis can be found in \cite{Braun:2010tt}. 

The ansatz for the effective action in an LPA'-type truncation reads
\begin{align}
\begin{split}
 \Gamma_k[\bar \psi,\psi,\sigma] = \int \mathrm{d}^d x \Bigg\{&  \bar \psi (Z_{\psi,k} \mathrm{i} \slashed{\partial} + \mathrm{i}\bar h_k \sigma)\psi \\
 &+ \frac{1}{2}Z_{\sigma,k} (\partial_\mu \sigma)^2 + U(\sigma^2)\Big\} \, .
 \end{split}
\end{align}
The bosonic ($Z_{\sigma,k}$) and fermionic ($Z_{\psi,k}$) wave function renormalisations, and the Yukawa coupling $\bar h_k$,
which transfers the interaction between bosons and fermions, are assumed to be scale-dependent
but field-independent. In the following studies, we again employ a Litim-type cutoff \cite{Litim:2001up}.

\subsection{Large-\texorpdfstring{$N$}{N} analysis}

The large-$N$ approximation is a good test case because the fixed point equations can be solved analytically.
Interestingly the scalar anomalous dimension does not vanish in contrast to the one of the Wilson-Fisher fixed point.
Even the fixed point potential looks very different.
The fixed point equations in the large-$N$ limit are given by \cite{Braun:2010tt}
\begin{align}
\begin{split}
 0=&(-2+\eta_\sigma) u'(\rho)+(d-2+\eta_\sigma) u''(\rho) \rho \\
 &+ \frac{8 \d_\gamma v_d}{d} \left(1-\frac{\eta_\psi}{d+1}\right)\frac{h^2}{(1+2 h^2 \rho)^2} \, ,
 \end{split}\\
 0=&(d-4+2\eta_\psi+\eta_\sigma)h^2 \label{eq:GN_LargeN_Yukawa} \, ,\\ 
 \eta_\sigma =& 8\frac{d_\gamma v_d}{d} h^2 \left( \frac{3}{4}+\frac{1-\eta_\psi}{2d-4} \right) \, , \\
 \eta_\psi =& 0,
\end{align}
again denoted in dimensionless quantities $u(\rho) = k^{-d}U(\sigma^2)/N$, $\rho = Z_{\sigma,k} k^{2-d}\sigma^2/(2N)$
and $h^2 =  Z_{\sigma,k}^{-1}Z_{\psi,k}^{-2}k^{d-4}\bar h^2 N$, and where $d_\gamma$ stands for the dimension of the Dirac algebra.
Note that for the large-$N$ limit, an appropriate rescaling has been taken into account.
\begin{figure}
\includegraphics[width=\columnwidth]{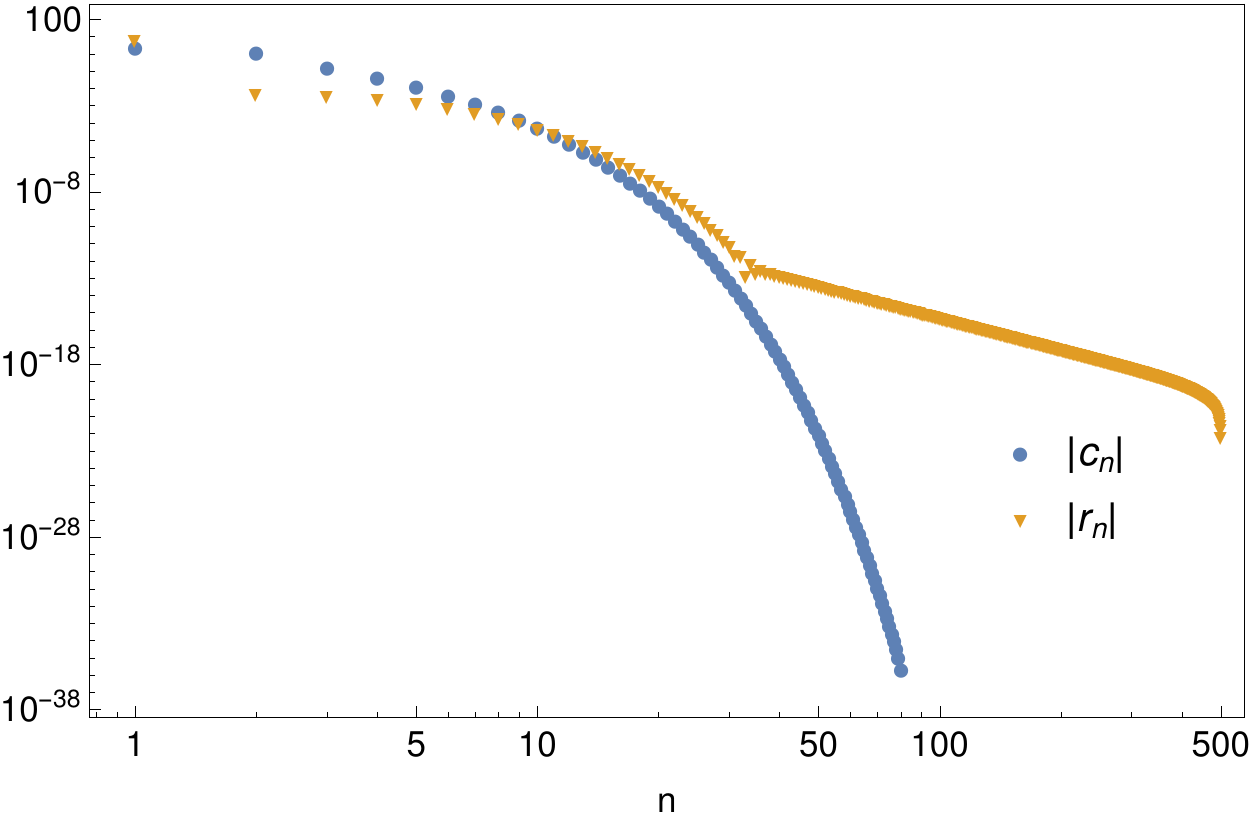}
\caption{Decay of coefficients of Chebyshev and rational Chebyshev expansion in the large-$N$ limit of the Gross-Neveu model.}
\label{fig:GN_largeN_coeff}
\end{figure}

In this approximation, we encounter a first order system.
The bosonic anomalous dimension can be read off from \eqref{eq:GN_LargeN_Yukawa} to be $\eta_\sigma = 1$ exactly (as long as $h \neq 0$). 
We can reproduce this result to all digits of {\tt float128} which gives an accuracy of about $10^{-32}$.
The exact fixed point value of the Yukawa coupling reads
\begin{equation}
 h_*^2 = \left( \frac{d}{d_\gamma v_d} \right) \frac{(d-4)(d-2)}{(8-6d)} \, ,
\end{equation}
and can be confirmed up to $10^{-32}$ as well.
The fixed point potential is given by the Gaussian hypergeometric function \cite{Braun:2010tt},
\begin{align}
\begin{split}
 u_*(\rho) = &-\frac{4(8-6d+d^2)}{3d-4}\rho\,\times\\
 &{}_2F_1\left(\frac{1}{1-d},1;\frac{2-d}{1-d};\frac{d}{d_\gamma v_d}\frac{8-6d+d^2}{3d-4}\rho\right).
 \end{split}
\end{align}
% %
% \begin{figure}
% \includegraphics[width=\columnwidth]{GN_LargeN_potdiff}
% \caption{Absolute difference of the analytic solution to the numerical solution in the large-$N$ limit of the Gross-Neveu model.}
% \label{fig:GN_largeN_pot}
% \end{figure}
% %

The absolute difference between the analytic solution and our numerical one 
% is depicted in figure \autoref{fig:GN_largeN_pot}.
% It shows that the error of our numerical solution 
can be estimated to be smaller than $3\times10^{-17}$ for large $\rho$.
For finite $\rho$ it is even smaller.
This is due to the Gaussian grid which only has points at finite $\rho$. Thus the asymptotic prefactor is only tuned regarding finite field values
and, therefore, has a larger error of about $3\times10^{-17}$.
For this calculation we have used $x_0 = 3/10$ and $L=2$.
The decay of the coefficients can be seen in \autoref{fig:GN_largeN_coeff}.
The Chebyshev expansion shows exponential convergence.
By contrast, the rational Chebyshev coefficients decrease exponentially at first, but only up to a certain number of coefficients,
the actual convergence rate is algebraic.
This is to be expected due to the asymptotic behaviour $\propto \sqrt{x}$.
The behaviour of the last coefficients shows a truncation effect which is not a numerical effect.
If we calculate the spectral coefficients from the analytic solution, we actually obtain a good agreement with the numerically calculated ones.
Ignoring the last coefficients affected by the truncation we read off $\sim 10^{-19}$ for the lowest coefficient.
The rule of thumb that the error can be estimated by $N\cdot c_N$
is in very good agreement with the maximal deviation of about $3\times10^{-17}$ from the exact solution.
\begin{figure}
\includegraphics[width=\columnwidth]{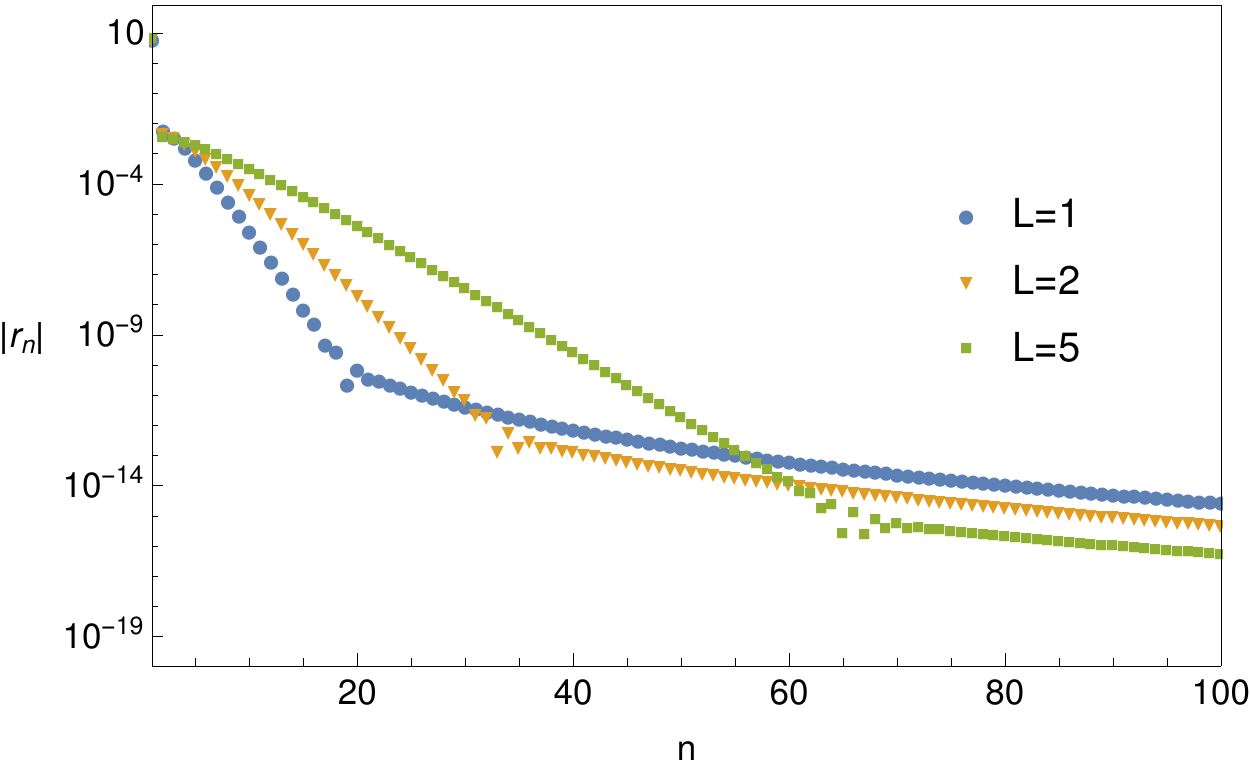}
\caption{Decay of rational Chebyshev coefficients in dependence on $L$ in the large-$N$ limit of the Gross-Neveu model.}
\label{fig:GN_largeN_changeL}
\end{figure}
\begin{figure*}
\includegraphics[width=\textwidth]{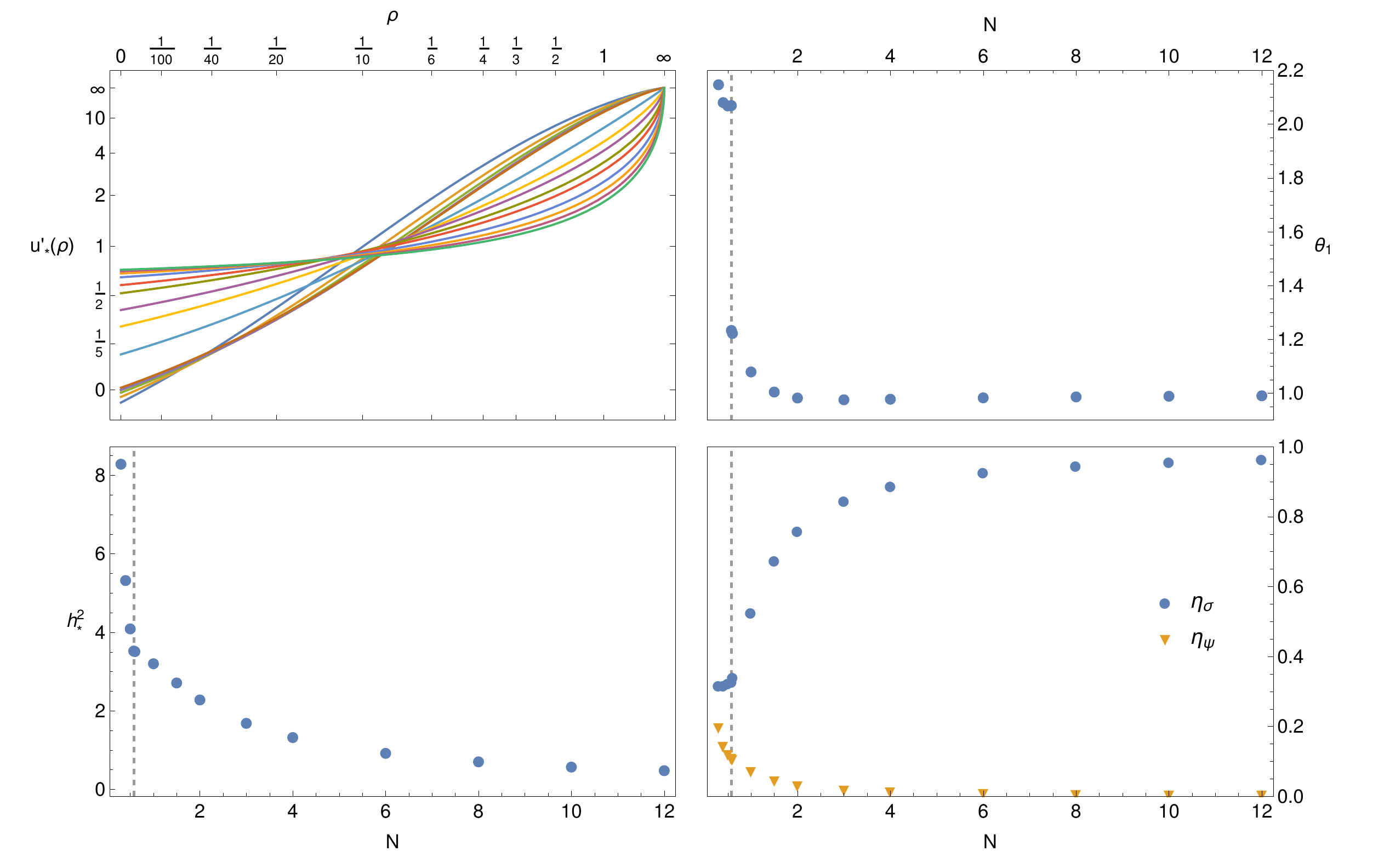}
\caption{Fixed point potential (upper left panel), relevant exponent (upper right panel), 
Yukawa-coupling (lower left panel) and anomalous dimensions (lower right panel) for the Gross-Neveu model with $0.3\leq N \leq 12$.
For increasing flavor number $u'_*(0)$ increases as well. The dashed line depicts the flavor number $N_t$ of the transition.}
\label{fig:GrossNeveu_finiteN}
\end{figure*}
Let us have a closer look at the choice of the two parameters $x_0$ and $L$.
They have quite some influence on the convergence behaviour of our coefficients.
We observe that if the matching point $x_0$ is chosen to be smaller, then the decrease of the Chebyshev coefficients is exponential as well, but faster.
This also holds for the rational coefficients if the matching point is increased.
However, lowering $x_0$ the algebraic convergence sets in earlier, enlarging $x_0$ the algebraic convergence sets in later.
It is remarkable that the gain of accuracy is inappreciable when taking enough coefficients into account 
because if algebraic convergence has set in, the coefficients do not change significantly.
This is different for the parameter $L$.
\autoref{fig:GN_largeN_changeL} shows that one can gain orders of magnitudes of accuracy if $L$ is increased.
The decrease of the first coefficients is more slowly, but the algebraic convergence sets in later.
This short analysis already makes clear that the choice of optimised parameters can strongly depend on the maximal number of coefficients that one takes into account.
These observations from one specific example may give an indication for other calculations as well.

\subsection{Finite \texorpdfstring{$N$}{N} analysis}

For finite $N$, the fixed point potential shows some interesting behaviour.
In \cite{Braun:2010tt} it is shown that the fixed point potential lies in the symmetric regime for all $N\geq2$.
For that reason, the potential was expanded polynomially.
A study of the convergence radius indicates the reliability of these results.
Unfortunately, the convergence is less clear for smaller $N$ such that a global solution is required.
For small $N$ the fixed point potential moves from the symmetric regime to the spontaneously symmetry-broken regime.
In \cite{Hofling:2002hj} a fixed point potential for $N=\frac{1}{2}$ (corresponding to one Dirac fermion
in the irreducible representation) was found in the symmetry-broken regime.
In \cite{Braun:2010tt} it was assumed 
that the non-Gaussian Gross-Neveu fixed point interpolates between the large-$N$ fixed point and the Wilson-Fisher fixed point in the $N\rightarrow0$ limit.

The fixed point equations for the Gross-Neveu model read, for general values of $N$ \cite{Gies:2013fua},
\begin{align}\allowdisplaybreaks
 0=&(-2+\eta_\sigma) u'(\rho)+(d-2+\eta_\sigma) u''(\rho) \rho \notag \\
 &- \frac{4 v_d}{d} \left(1-\frac{\eta_\sigma}{d+2}\right) \frac{3 u''(\rho)+2\rho u'''(\rho)}{(1+u'(\rho)+2 \rho u''(\rho))^2} \notag \\
 &+ \frac{8 \d_\gamma v_d}{d} N \left(1-\frac{\eta_\psi}{d+1}\right)\frac{h^2}{(1+2 h^2 \rho)^2} \, , \\
 0=&(d-4+2\eta_\psi+\eta_\sigma)h^2 \notag \\
 &+\frac{16 v_d}{d} h^4 \left(\frac{1-\frac{\eta_\psi}{d+1}}{1+2\rho_0 h^2}+\frac{1-\frac{\eta_\sigma}{d+2}}{1+u'(\rho_0)+2\rho_0 u''(\rho_0)}\right)\times \notag \\
 &(1+2\rho_0 h^2)^{-1}(1+u'(\rho_0)+2\rho_0 u''(\rho_0))^{-1}\notag \\
 &-\frac{2 v_d}{d} h^4 \left(48\rho_0 u''(\rho_0)+32 \rho_0^2 u'''(\rho_0) \right) \times \notag \\
 &\left(\frac{1-\frac{\eta_\psi}{d+1}}{1+2\rho_0 h^2}+2 \frac{1-\frac{\eta_\sigma}{d+2}}{1+u'(\rho_0)+2\rho_0 u''(\rho_0)}\right)\times \notag \\
 &(1+2\rho_0 h^2)^{-1}(1+u'(\rho_0)+2\rho_0 u''(\rho_0))^{-2}\notag \\
 -&\frac{64 v_d}{d} h^6 \rho_0 \left(2 \frac{1-\frac{\eta_\psi}{d+1}}{1+2\rho_0 h^2}+\frac{1-\frac{\eta_\sigma}{d+2}}{1+u'(\rho_0)+2\rho_0 u''(\rho_0)}\right)\times \notag \\
 &(1+2\rho_0 h^2)^{-2}(1+u'(\rho_0)+2\rho_0 u''(\rho_0))^{-1} \, , \\
 \eta_\sigma =& 8\frac{d_\gamma v_d}{d} h^2 N \times \notag \\
 &\left( \frac{1-2\rho_0 h^2}{(1+2\rho_0 h^2)^4}+\frac{\frac{1}{4}+\frac{1-\eta_\psi}{2d-4}}{(1+2\rho_0 h^2)^2} 
 +\frac{\frac{1-\eta_\psi}{d-2}}{(1+2\rho_0 h^2)^3}\right)\notag \\
 &+\frac{8 v_d}{d}\rho_0\frac{(3u''(\rho_0)+2\rho_0 u'''(\rho_0))^2}{(1+u'(\rho_0)+2\rho_0 u''(\rho_0))^4} \, , \\
 \eta_\psi =& \frac{8 v_d h^2}{d}\frac{ 1-\frac{\eta_\sigma}{d+1}}{(1+2\rho_0 h^2)(1+u'(\rho_0)+2\rho_0 u''(\rho_0))^{2}},
\end{align}
with $u(\rho) = k^{-d}U(\sigma^2)$, $\rho = Z_{\sigma,k} k^{2-d}\sigma^2/2$ and $h^2 =  Z_{\sigma,k}^{-1}Z_{\psi,k}^{-2}k^{d-4}\bar h^2$.
The asymptotic behaviour of the potential is given by $u'(\rho) \propto \rho^{\frac{2-\eta_\sigma}{d-2+\eta_\sigma}}$.

\begin{figure*}
\includegraphics[width=\textwidth]{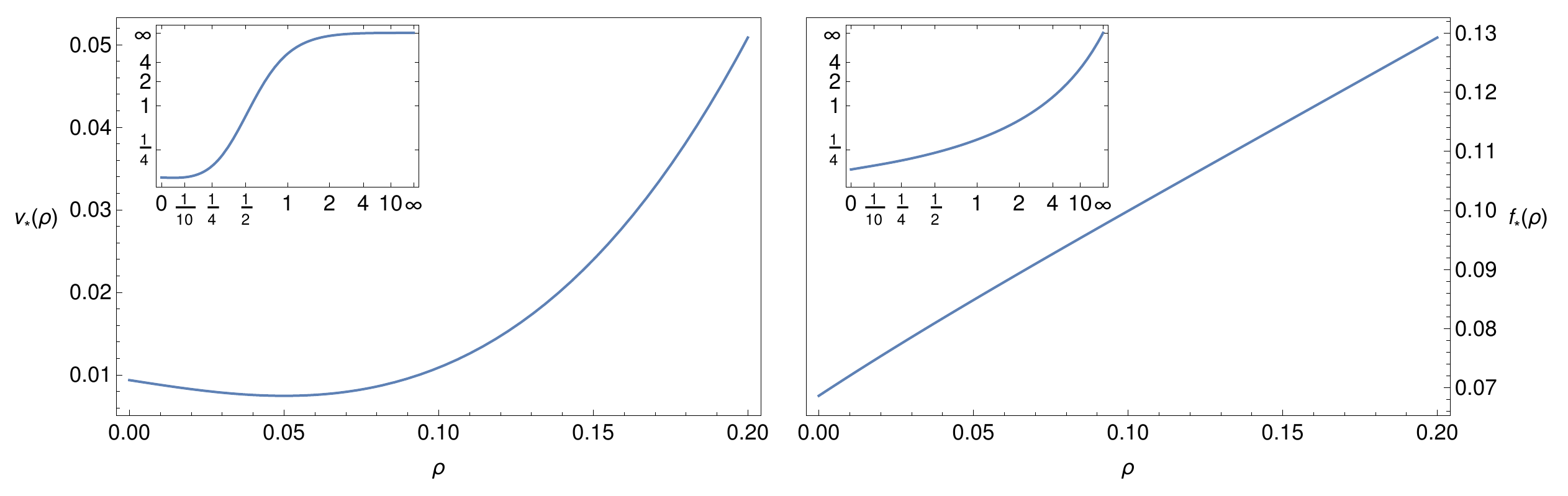}
\caption{Global fixed point solution for the scalar-tensor model in three dimensions. The scalar potential
closely resembles the Wilson-Fisher fixed point potential on flat background. The non-minimal coupling is strictly positive, admitting a positive
Newton constant.}
\label{fig:scalar_tensor_sol}
\end{figure*}

We have calculated the global solution to these fixed point equations and the relevant critical
exponent in three dimensions for $N$ lying between $0.3$ and $12$, see \autoref{fig:GrossNeveu_finiteN}.
We obtain a very good agreement with \cite{Braun:2010tt} for $N\geq2$ where a polynomial approximation is employed.
Even the relevant exponent matches in the first four relevant digits.
It is worth mentioning that this good agreement is only obtained by taking high orders in the polynomial truncation into account \cite{Braun:2010tt}, especially for small $N$.
Our results for $N\geq2$ are also compatible with other methods such as $1/N$-expansions \cite{Gracey:1993kc,Vasiliev:1992wr},
and Monte-Carlo simulations \cite{Hands:1992be,Karkkainen:1993ef}.
In fact, systematic truncation errors appear to be smaller for the Gross-Neveu model in comparison with the $O(1)$ model.
The overall consensus among the non-perturbative methods is very satisfactory.

Let us now concentrate on the small-$N$ regime.
The transition from the symmetric to the symmetry-broken regime can be determined to be at $N_t\approx 0.5766$.
As a new result, we observe that the Gross-Neveu fixed point does not approach the Wilson-Fisher fixed point for small $N$.
This can be seen from the behaviour of the Yukawa coupling and the anomalous dimensions on the one hand and the relevant exponent on the other hand.
In particular, the behaviour of $h_*$ suggests that the Gross-Neveu fixed point moves to infinity in theory space for $N\to0$.

It is instructive to compare our results for $N=1/2$ with those of \cite{Hofling:2002hj} where also a full potential flow has been studied (note that
our convention of $N=1/2$ corresponds to $N=1$ in \cite{Hofling:2002hj}; for aspects of criticality see \cite{Janssen:2012pq}).
In \cite{Hofling:2002hj} the fluctuation terms $\propto \rho_0 u''(\rho_0)h^4$, $\propto \rho_0^2 u'''(\rho_0)h^4$ and $\propto \rho_0 h^6$ have been missed
in the derivation of the flow equation, see the discussion in \cite{Gneiting:2005}.
If we artificially switch off these terms, the vacuum expectation value and the critical exponent of our calculation are in good agreement with those of \cite{Hofling:2002hj}.
On the contrary, including these terms, even the first relevant digit changes.
For $N=1/2$, we obtain $\nu = 1/\theta_1 = 0.4836$, $\eta_\sigma = 0.3227$, $\eta_\psi = 0.1204$.
In conclusion it is remarkable that our approach is able to find a global solution in a regime where a polynomial truncation is not reliable.

\section{Scalar-tensor gravity}\label{sec:STG}

As a final example, let us consider a model which couples a scalar field non-minimally to gravity \cite{Narain:2009fy,
Henz:2013oxa,Percacci:2015wwa}.
The ansatz for the effective average action is given by
\begin{align}
 \Gamma_k[\sigma,g] &= \int \mathrm{d}^dx \sqrt{g} \left( V_k(\sigma) - F_k(\sigma) R + 
 \frac{1}{2} g^{\mu\nu}\partial_\mu\sigma\partial_\nu\sigma \right) \, ,
\end{align}
where $g$ denotes the determinant of the metric $g_{\mu\nu}$ and $R$ is the Ricci scalar. It serves as an effective model for the cosmological evolution of
the Universe. Here, we want to use the flow equations of \cite{Percacci:2015wwa}, in which an exponential split was used to quantise the
gravitational fluctuations.
The explicit flows of the dimensionless variants $v(\rho)$ and $f(\rho)$ of the functions $V(\sigma)$ and $F(\sigma)$ can be looked
up in \cite{Percacci:2015wwa}.

\begin{figure*}
\includegraphics[width=\textwidth]{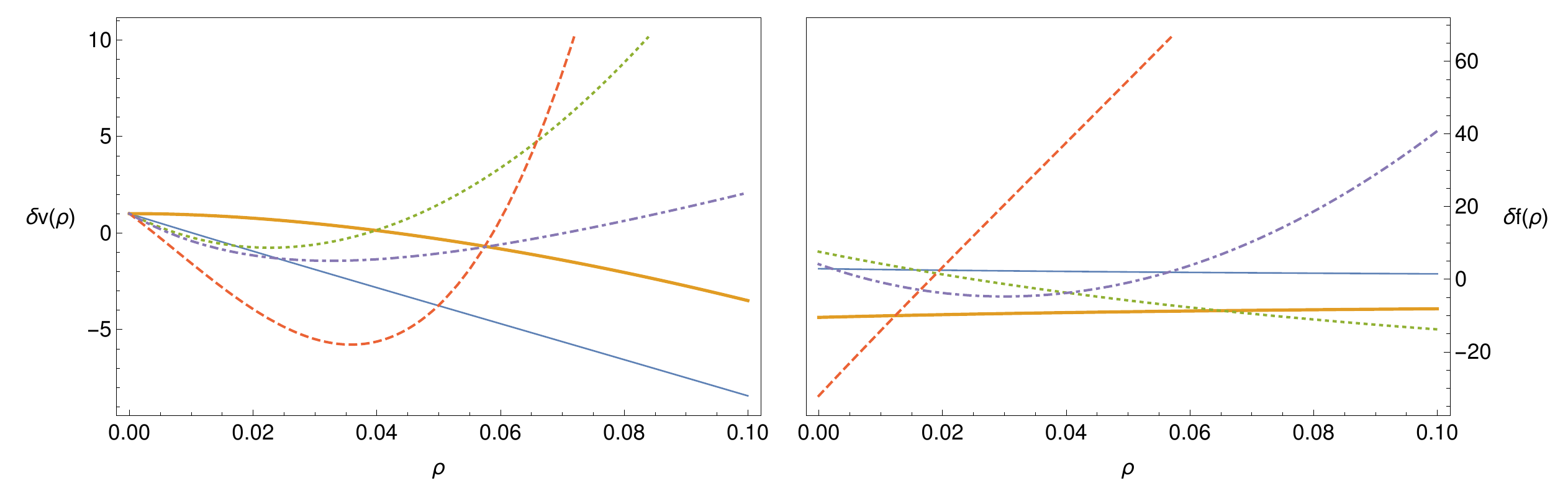}
\caption{Eigenfunctions of the fixed point solution to the three-dimensional scalar-tensor model. They are normalised such that
$\delta v(0)=1$. From relevant to irrelevant critical exponents: thin blue, thick orange, dotted green, dashed red, dot-dashed violet.}
\label{fig:scalar_tensor_eig}
\end{figure*}

We have found a non-trivial solution to the fixed point equations, which is shown in \autoref{fig:scalar_tensor_sol}.
Some remarks are in order. First, the effective potential of
the scalar field closely resembles the Wilson-Fisher fixed point potential on flat background. Second, note that $f_*(\rho)>0$ for all $\rho\geq0$,
which implies that the (analogue of the) Newton constant is positive, and we are indeed in the physical regime. Third, the minimum
of the potential $v(\rho)$ lies at $\rho = 0.05004$.
Lastly, we observe exponential convergence of the expansion.
Thus, the solution can in principle be computed to arbitrary precision.

Let us now turn our attention to the critical exponents of the fixed point. The six leading critical exponents are
\begin{align}
 \theta_1 &= 3 \, , \qquad &\theta_2 &= 1.9134 \, , \notag \\
 \theta_3 &= 1.1798 \, , \qquad &\theta_4 &= 0.6679 \, , \\
 \theta_5 &= -0.2812 \, , \qquad &\theta_6 &= -1.217 \, . \notag
\end{align}
Notice that the first exponent is present in any fixed point solution. It is related to what one would call cosmological constant, being
the most relevant perturbation. The eigenfunctions for all of them, except for the first one, are given
in \autoref{fig:scalar_tensor_eig}. For the first critical exponent $\theta_1$, the eigenfunction is $(\delta v=1,\delta f=0)$.

Let us compare these findings with those of \cite{Percacci:2015wwa}. There, the fixed point equations were first analysed in a one-loop approximation.
Using a local field expansion, two non-trivial solutions have been identified: the conformal solution $v = 1/(18\pi^2), f = 37/(72\pi^2)+\sigma^2/4$, possessing
four relevant directions, and a Wilson-Fisher-like solution with three relevant directions. There are indications that the latter fixed point runs into a singularity
due to $f$ having a zero. On the basis of the full equations, no conformal solution with the same simple structure as in one-loop approximation was found.
No statement was given on the fate of the Wilson-Fisher-like solution.

Our results are the following: the critical exponents of our global solution are very close to the ones of the conformal solution in the one-loop approximation.
Also, the general form of the non-minimal coupling $f(\rho)$ is qualitatively the same.
On the other hand, the form of the scalar potential $v(\rho)$ is Wilson-Fisher-like.

We further checked deviations from three dimensions. For any dimension between three and four, we find a similar fixed point solution.
Though, at $d=4$ exactly, this solution seems not to be present anymore, reminiscent of the situation in a purely scalar theory.
We take this as an indication that the solution found here might be the generalisation of the Wilson-Fisher fixed point.

Finally, let us comment on why solutions, where the non-minimal coupling becomes negative, $f(\rho)<0$, should be taken with care.
In the derivation of the flow equation, the regulator was spectrally adjusted. In particular, for the tensor fluctuations, the regulator was chosen to be
proportional to $f(\rho)$. As soon as this function crosses zero, the tensor modes are not properly regularised anymore.
In the flow equations, this is reflected by terms proportional to $1/f(\rho)$, which can only be compensated by divergences in derivatives of the potential.
This might explain why we could not find further global solutions.

Summarising, we find a globally well-defined solution in scalar-tensor gravity. However, the inclusion of matter,
in particular scalar fields, induces higher order curvature terms. Thus, one should check that the global
solution survives. On the other hand, for cosmological applications, an Einstein-Hilbert type of truncation
as discussed here is expected to be well-suited as an effective model.

\section{Conclusions}

In this work, we presented a method to solve ODEs globally.
Pseudo-spectral techniques are not new and already have been applied to various problems in physics.
A lot of rigorous results on spectral or pseudo-spectral methods are available \cite{Boyd:ChebyFourier}.
However, to our knowledge the expansion in rational Chebyshev polynomials is not very well established in quantum field theory calculations,
although it allows to investigate the question of global existence of solutions.
This is a very important question since the non-linear ODEs encountered in FRG
studies can have many more or less stable local solutions
which are not easy to distinguish from global ones if only local information is accessible.
For instance, the physical criteria of polynomial boundedness and self-similarity are difficult to impose locally \cite{Morris:1998da}.
The method presented in this paper offers a comparatively easy way to find global solutions of ODEs that circumvents such pitfalls.

We applied this method to various models.
The first test case was the very well known $O(1)$ model in three dimensions which we considered in both the LPA and LPA' truncation.
There are numerous works on expansions for small and large fields and results gained via the shooting method
which give a good impression of the global behaviour of the  Wilson-Fisher fixed point.
We reported on the difference between LPA and LPA' truncations taking the global behaviour of the potential into account.
Although the anomalous dimension is very small, the asymptotic behaviour, especially with regard to the prefactor, changes significantly.
Besides the fixed point potential itself, we calculated the eigenfunctions globally and determined the critical exponents.
For all quantities we obtained good agreement with already known results calculated with other methods.
As far as numerical accuracy is concerned, our method outperforms previous results by many orders of magnitude, while being
very stable, fast and lightweight.

Subsequently, we extended our study to fractional dimensions, taking $d=2.4$ as a representative.
We found all multi-critical fixed point potentials predicted in \cite{Codello:2012ec}
and could, moreover, determine their global behaviour.
We were able to see the next higher critical fixed point emerging at $d<2.4$ which demonstrates
that our numerical method is highly accurate and stable.
All physical quantities, the anomalous dimension and critical exponents, again match with earlier results.

As a next system we considered the Gross-Neveu model. 
On the one hand, the large-$N$ limit provides an easily accessible analytic solution.
On the other hand, the small-$N$ limit is not easily accessible by use of common local expansions
and offers, therefore, the possibility to demonstrate the advantages of our global method. 
In the large-$N$ case we obtained a conclusive agreement with analytic results.
For finite $N$, our results agree very well with other data, including $1/N$-expansions and lattice methods.
We were able to fix the transition flavor number to be $N_t\approx0.5766$
and observe how the fixed point potential goes over from the symmetric to the symmetry-broken regime.
We found that the fixed point Yukawa coupling grows large for $N\to0$.
The anomalous dimensions, in particular the fermionic one, take on finite values.
This suggests that the Gross-Neveu fixed point does not merge with the Wilson-Fisher fixed point in the limit $N\to0$ contrary
to what has been anticipated in \cite{Braun:2010tt}.
Additionally, we saw that all fluctuation terms in the Yukawa fixed point equation which occur in the symmetry-broken regime
have a significant influence on physical quantities, such as critical exponents.
Comparing to \cite{Hofling:2002hj} where some fluctuation terms were missed we determine the deviation to be up to $30\%$.

We finally discussed a scalar-tensor gravity model in $d=3$.
This model is supposed to have a gravitationally dressed Wilson-Fisher fixed point.
We found a global solution implying a positive Newton constant with four relevant critical exponents.
No other non-trivial global solution was found, though it may be difficult to prove that our solution is unique.
We could recover the solution in all dimensions smaller than four, but not in $d=4$. This is taken as an indication
that the solution found might be the generalisation of the Wilson-Fisher fixed point to curved space.

We emphasise that the models treated in this paper only represent a small set of examples for a large amount of possible applications of pseudo-spectral methods.
These methods are straightforwardly extendable to include more than one variable which is needed if additional invariants are taken into account.
Full potential flows requiring the technique to be extended to PDEs are underway.

\section*{Acknowledgements}

We would like to thank Marcus Ansorg, Alexander Blinne, Holger Gies and Andreas Wipf for useful discussions,
and Holger Gies and Andreas Wipf for comments on the manuscript.
This work was supported by the DFG-Research Training Group
``Quantum- and Gravitational Fields'' GRK 1523/2.
JB acknowledges further support by DFG under grant no. Gi 328/6-2 (FOR 723), while
BK acknowledges funding by DFG grant no. Wi 777/11-1.
Finally, we would like to thank for the positive feedback we received on the
7th International Conference on the Exact Renormalization Group (ERG 2014).

% \bibliography{specbib}
%

\end{document}